\documentclass[11pt,preprint]{aastex}
\usepackage{amsmath}
\usepackage{graphics}
\usepackage{epstopdf}
\newcommand{\rocke}{ROCKE-3D}

\begin{document}
\title{Climates of Warm Earth-like Planets III: Fractional Habitability from a Water Cycle Perspective}
\author{Anthony D. Del Genio, M.J. Way\altaffilmark{1,2}, Nancy Y. Kiang\altaffilmark{2}, Igor Aleinov\altaffilmark{3}, Michael J. Puma\altaffilmark{3}, Benjamin Cook}
\affil{NASA Goddard Institute for Space Studies, 2880  Broadway, New York, NY, 10025, USA}
\email{anthony.d.delgenio@nasa.gov}

\altaffiltext{1}{Theoretical Astrophysics, Department of Physics and Astronomy, Uppsala University, Uppsala, SE-75120, Sweden}
\altaffiltext{2}{GSFC Sellers Exoplanet Environments Collaboration}
\altaffiltext{3}{Center for Climate Systems Research, Columbia University, 2880  Broadway, New York, NY, 10025, USA}

\begin{abstract}
The habitable fraction of a planet's surface is important for the detectability of surface biosignatures.  The extent and distribution of habitable areas is influenced by external parameters that control the planet's climate, atmospheric circulation, and hydrological cycle. We explore these issues using the \rocke{} General Circulation Model, focusing on terrestrial water fluxes and thus the potential for the existence of complex life on land. Habitability is examined as a function of insolation and planet rotation for an Earth-like world with zero obliquity and eccentricity orbiting the Sun. We assess fractional habitability using an aridity index that measures the net supply of water to the land. Earth-like planets become ``superhabitable'' (a larger habitable surface area than Earth) as insolation  and day-length increase because their climates become more equable, reminiscent of past warm periods on Earth when complex life was abundant and widespread. The most slowly rotating, most highly irradiated planets, though, occupy a hydrological regime unlike any on Earth, with extremely warm, humid conditions at high latitudes but little rain and subsurface water storage.  Clouds increasingly obscure the surface as insolation increases, but visibility improves for modest increases in rotation period.  Thus, moderately slowly rotating rocky planets with insolation near or somewhat greater than modern Earth's appear to be promising targets for surface characterization by a future direct imaging mission.  
\end{abstract}
\section{Introduction}
The stability of liquid water on a planet's surface is considered a necessity for remotely detectable Earth-like life \citep{DesMarais2002,Schwieterman2018}. This requirement constrains traditional 1-dimensional definitions of the ``habitable zone" of distances from the host star that define the requirements for a future direct imaging exoplanet mission that will search for life \citep{Kane2016}. However, whether a planet within the habitable zone actually is habitable depends on more factors than just planet-star distance, and likewise, planets outside the traditional habitable zone may be habitable.

Furthermore, although habitability is sometimes defined to be a binary assessment for a planet \citep{Cockell2016}, climate varies over a planet's surface. Inhabited planets are likely to range from those with abundant, widespread life to those with only small niches for life. Future spacecraft missions and ground-based observatories will undoubtedly discover many more potentially habitable planets than can be observed in detail for evidence of biosignatures. We assume that more abundant life increases the odds that biosignatures can be detected on a planet's surface (where ``abundant" can have different metrics:  surface cover and standing biomass, both of which are not necessarily correlated with net productivity). This is exemplified by many Earthshine and Earth simulator studies that quantify detectability for different amounts of surface cover by vegetation by planetary phase \citep{Seager2005,Montanes2006,Tinetti2006,Arnold2008}. It is therefore worthwhile to consider which types of planets produce land surface conditions that allow for the greatest extent of habitable climates.  We assume for this study that life forms consistent with the terrestrial Plant Kingdom classification may evolve on habitable exoplanets and thus we look for environments likely to support extensive vegetative life, though this may not be the case.  In support of this assumption, we note that \citet{CockellMBoC2016} argues that life forms on Earth have universal characteristics that we expect will emerge elsewhere in the universe.

The areal extent of exoplanet habitability has been considered from several different viewpoints. \cite{Spiegel2008} used a 1-dimensional energy balance model to consider how planet rotation and land-ocean fraction affect the fraction of the planet surface that lies within a temperature range conducive to life. They coined the term ``fractional habitability" as a more continuous descriptor than a simple yes/no characterization of a planet's habitability. Similar definitions have been used in several other studies \citep[e.g.][]{Silva2017,Wolf2017,Jansen2019}. Numerous studies have examined the effects of obliquity on the latitudinal distribution of above-freezing conditions, including how this differs for all-land and all-ocean planets \citep{Abe2003,Kilic2018,Colose2019}.  \cite{HA2014} discussed properties that might make a planet ``superhabitable," i.e., having a greater habitable surface area than Earth: A planet area (volume) greater than Earth's, an optimal distribution of land and ocean, a stellar habitable zone wider than Earth's, or a longer duration of the continuous habitable zone associated with prolonged plate tectonics, other greenhouse gas recycling mechanisms, and magnetic shielding. \cite{Schwieterman2019}, on the other hand, noted that complex life as we know it is not suited to the high CO$_2$ and CO abundances expected near the outer edge of the habitable zone and for planets orbiting M stars, respectively.

Similar considerations arise in the history of the terrestrial planets in our Solar System.  Earth has been continuously habitable since the late Hadean or early Archean, but it has experienced partial or complete ``snowball" periods when much or all of its surface was covered by ice, although life still survived within the ocean \citep{SohlChandler2007}. At other times, Earth was warmer than today, with clement conditions due to increased CO$_{2}$ concentrations and extensive surface life extending to the poles.  These ``equable" climate periods \citep{Lunt2012,Chandler2013} may be thought of as examples of ``superhabitable" planets. Mars today is probably uninhabitable, but geologic evidence indicates that liquid water was present in its ancient past. Whether Mars ever had high enough temperatures to sustain a primordial northern ocean and widespread habitability for an extended period of time \citep[e.g.][]{Rodriguez2015,Villanueva2015} or only episodic, localized clement conditions and liquid water \citep[e.g.][]{Wordsworth2016,Kite2017} is still unclear. In any case, with very high CO$_2$ amounts only primitive life would have been possible.  Venus today is far too hot to sustain life and has lost almost all its water, but scenarios exist by which a shallow ocean consistent with isotopic constraints might have maintained habitable temperatures \citep{Way2016}.   

Although surface liquid water is a precondition for detectable exoplanet life, only a few studies have paid attention to water itself as a metric for habitability despite the ``follow the water" mantra of astrobiology.  Regional water availability is the controlling factor for the distribution of land-based life on Earth. For example, satellite observations of Earth demonstrate that an index of vegetation based on the same ``red edge" concept that has been suggested as an exoplanet biosignature is more sensitive to water availability over a larger fraction of the Earth's land surface than to either temperature or cloud cover (a proxy for surface insolation) \citep{Seddon2016}.  \cite{Abe2011} showed that on an all-land Earth-like planet with a limited subsurface water reservoir, the tropics would dry and water would collect at the poles. Titan is too cold to have liquid water, but methane exists in liquid form in lakes on its surface, preferentially at the poles. \cite{Mitchell2008} showed that this is the combined result of Titan's slow rotation, seasonality, the dynamics of a moist atmosphere, and the depth of the subsurface methane reservoir. \cite{YangLiuHuAbbot2014} considered the conditions under which water on a synchronously rotating planet with an ocean and exposed land would be trapped as land ice on the nightside, rendering the planet uninhabitable.   

It is therefore useful to ask which factors determine the distribution of water over an exoplanet's land surface. Three-dimensional general circulation models (GCMs) are ideal tools to address such questions, because they predict the locations and extent of regions of upwelling and downwelling that determine rainfall and evaporation patterns and thus the exchange of stored water with the atmosphere. We are interested in quantifying the heterogeneous distribution of liquid water over land surfaces, because land-based life forms can produce surface signatures of extant life from space \citep[e.g.][]{DesMarais2002,Arnold2008} as well as volatiles that are potential biosignatures \citep{Seager2016}.

Worlds with all-ocean surfaces (aquaplanets) are of course habitable by the surface liquid water definition. However, they have several challenges in supporting abundant life: A narrower habitable zone than planets with land \citep{Abe2011}; possible difficulties in supporting a carbonate-silicate cycle feedback to stabilize the climate \citep{Abbot2012}; an unstable CO$_{2}$ cycle due to temperature-dependent CO$_{2}$ dissolution \citep{Kitzmann2015}; and sparseness of nutrients (oligotrophy) in open oceans \citep{Behren2005}. In situ signatures of extant marine or aquatic biota such as algae would likely be too sparse, ephemeral, or localized to be useful for telescope observation. It may also be more difficult to detect biosignatures on aquaplanets due to enhanced abiotic signals \citep{Desch2017}.  GCMs are also not well-suited to studying ocean biosignatures, because the most productive aquatic environments are in shallow coastal zones that a GCM cannot spatially resolve. Finally, since GCMs are geared to simulating climate on century or shorter time scales, they are not appropriate for exploring geologic time scale accumulation of biogenic gases such as oxygen.  The potential of alternative biogenic gases from ocean biota to be detectable as atmospheric biosignatures remains an area of investigation \citep{Schwieterman2018}.

In this paper, we therefore focus on the question of land surface water availability with the \rocke{} GCM \citep{Way2017}.  An exhaustive study of all the parameters that can affect planetary climate and water distribution is beyond the scope of this work.  We choose instead to illustrate the idea by examining the behavior of a water metric, the aridity index, in the context of simulations of an Earth-like planet on which two parameters, insolation and rotation period, are varied. These parameters have previously been shown to have large effects on exoplanet climates \citep{Yang2013,Yang2014,Kopparapu2016, Way2018}.  This paper is Part III of a three-part series.  Part I \citep{Way2018} describes details of the global mean climate properties of planets at different insolations and rotation periods. Part II \citep{Jansen2019} examines the fractional habitability of these planets from a purely temperature standpoint as well as the effect of insolation on the carbonate-silicate cycle feedback. We consider a planet with Earth's land-ocean distribution as a useful, demonstrably habitable, template. What land fraction and distribution optimizes habitability and biosignature detection is a subject for future study. 

In Section 2 we describe the methods used in this study. Section 3 explores the dependence of habitable land surface area on insolation and rotation and the physical mechanisms that control it. Section 4 relates these results to biomes, by analogy to Earth, discusses the possibility of superhabitable planets as defined by the water cycle, and analyzes a simulated climate type with no Earth counterpart.  Section 5 discusses the limitations of our study and assesses the effects of clouds on detectability of surface features.  Our conclusions are presented in Section 6.

\section{Methods}\label{sec:Methods}
\subsection{Experimental Setup}\label{sec:Experimental_Setup}

The simulations used herein are detailed in \citet{Way2018},(hereafter, Paper I). In summary, we use the Planet\_1.0 version of  \rocke{}  described in detail in \cite{Way2017}. We run \rocke{} at 4$\degr$ $\times$ 5$\degr$ latitude $\times$ longitude resolution with 40 atmospheric layers and a top at 0.1 hPa. Surface air pressure is set to 984mb. The atmosphere is N$_{2}$ dominated with 400 ppmv CO$_{2}$ and 1 ppmv CH$_{4}$, i.e., modern Earth-like but without O$_2$, O$_3$, or aerosols. All simulations use a dynamic fully coupled ocean with the same horizontal resolution as the atmosphere. The ocean is bathtub-style with continental shelves set at a depth of 591 meters. The rest of the ocean has a depth of 1360 m. Land is treated as bare soil with 3.5 m depth, a dry albedo of 0.2, and made of a 50/50\% clay/sand mix for soil texture.  These physical properties limit available water at saturation in the soil column to 1.3229 m.  The texture is intermediate between the extremes of pure sand and pure silt, which are less and more efficient at retaining water, respectively, and conversely, more or less efficient at infiltration of water.  The continental distribution and topography are approximately that of modern Earth.

The baseline simulation has an incident solar flux at the substellar point of 1360.67 W/m$^2$, identical to Earth's modern solar constant, and a rotation period P = 1 d.  Obliquity and eccentricity are set to zero, so compared to modern Earth, our baseline has no seasonality and thus warmer/colder conditions at the equator/pole.  Relative to this baseline, we conduct a series of simulations that vary the incident sunlight relative to modern Earth (S0X) in steps of 0.1 and the rotation period in successive factors of 2 relative to Earth's rotation period. At each rotation period the highest S0X simulated is that for which the model is able to reach radiative balance with temperatures below 400 K, above which the radiation tables begin to lose accuracy and water vapor begins to become a non-negligible fraction of atmospheric mass.

\subsection{Aridity as a water cycle metric}

Aridity indices are commonly used to provide climate-based indicators of the regional availability of soil moisture for agriculture or to classify biomes and assess potential effects of anthropogenic climate change \citep[e.g.][]{Feng2013,Fu2014,SF2015,Fu2016}.  An aridity index measures the competition between the atmospheric supply of water to the surface (precipitation) and the evaporative demand of water from the surface.  In other words, it is a measure of the net water balance on land. 

An aridity index is appealing for studies of the climatic properties of Earth's land surface because it is readily available from time series of meteorological measurements. For exoplanets, its advantage is its utilization of atmospheric variables controlled largely by general circulation features predicted from external parameters that can be measured now (insolation) or may be measurable in the future (rotation period), as opposed to a more direct hydrological metric such as soil moisture that cannot easily be constrained for exoplanets or modeled as a measurable quantity in GCMs.  It can be considered a flux-based index of the hydrological cycle, as opposed to soil moisture, which is a reservoir-based index.  

\citet{Scheff2017} showed that an aridity index successfully captures patterns of soil moisture change in both a warmer 21st Century Earth climate and a colder Last Glacial Maximum climate relative to modern Earth.  Its weaknesses are its sensitivity to integration period (e.g., a complete Earth orbit vs. seasonal vs. the time scales of individual weather events, which can cause aridity to vary by 10-30\% according to \citet{SF2015}) and its inability to capture the effects of varying soil texture, depth, and runoff, which affect the actual soil moisture available for life. These can sometimes lead to unreliable results, e.g., when precipitation exceeds the infiltration capacity of the soil leading to runoff, when precipitation and evaporative demand or temperature are temporally out of phase, and when soil water availability varies temporally with freezing and thawing periods. None of these are expected to qualitatively affect the broad estimates of habitability that are the subject of this paper.

Evaporative demand is defined by the potential evapotranspiration (PET), the evapotranspiration rate that would occur given a well-watered surface. Many different expressions for PET exist: assuming an open water surface \citep{Penman1948}; incorporating a parameter to account for enhanced evaporation from taller vegetation canopies \citep{PT1972}; incorporating a stomatal resistance or conductivity for control of transpiration by vegetation \citep{Monteith1975}; and full soil water balances including storage and runoff \citep{Thornthwaite1948,KF1994}. For evaluating climate for vegetation, it is common to use the Penman-Monteith equation for PET, which estimates the rate at which water would evaporate from a wet surface (covered by well-watered vegetation) in energy balance with the atmosphere for a given climate state. We use a version of the Penman-Monteith equation described by \citet{Cook2014}, who use the standardized reference equation of evapotranspiration for the American Society of Civil Engineers and the Food and Agriculture Organization \citep{Allen1998,Allen2005}.  It is physically-based, incorporating information on wind speed, atmospheric humidity, vapor pressure deficits, and net radiation.  This differentiates it from more empirically-based approaches that depend only on temperature. The version we use is based on agricultural applications, assuming ground covered by a "reference crop" with stomatal resistance that limits transpiration.  Higher PET values would result from ignoring stomatal resistance, as would be appropriate for the bare soil planet we assume, but \citet{SF2014} show that this has a small effect relative to the primary climate controls on PET, which are our concern in this paper.  This equation also allows for comparisons to Earth climatology as a reference.  Other approaches to calculating PET also exist, e.g., the Priestley-Taylor equation \citep{PT1972}, but this makes an arbitrary assumption about the state of air flowing over the surface and thus introduces its own uncertainties.  \citet{Stev2019} recently estimated the aridity index distribution for a synchronously rotating Proxima Centauri b using an even simpler PET expression that simply scales linearly with temperature. 

We use the \emph{aridity index} (A) defined in terms of model-estimated PET and model precipitation (Pr) as in \citet{SF2015}
\begin{equation}
A = Pr/(Pr+PET)
\end{equation}   

This version of the aridity index is derived from the standard expression A = Pr/PET used in Earth climate studies, but the appeal of the \cite{SF2015} version we use is that it is bounded (unlike the standard expression), ranging from 0 to 1.  Thus fluxes into and out of wet soil are balanced when A = 0.5.   We focus on two subsets of A values corresponding to two extreme hydroclimatic regimes \citep{SF2015}:
\begin{itemize}
\item Humid climates: A $>$ 0.39
\item Arid/hyper-arid climates: A $<$ 0.17
\end{itemize}

These thresholds are based on terrestrial wetland/dryland categories defined by \cite{MiddletonThomas1997} and used by the United Nations Environmental Program. A $>$ 0.39 is typical of the eastern half of the United States, tropical rain forests, and boreal forests, while A $<$ 0.17 delineates most of the Earth's deserts \citep{SF2015}. In the exoplanet context, the precise values of these thresholds are somewhat arbitrary, so these Earth-based values are merely convenient points-of-reference.  Also note that a high aridity index is actually wet rather than arid, whereas a low value is dry.  Thus despite being named an ``aridity index", it is more intuitive to interpret A as a wetness index. Large values of A (Pr$>>$PET) indicate regions where there is sufficient water but evapotranspiration is energy-limited, while small values (Pr$<<$PET) occur in regions that lack sufficient water to reach the potential evapotranspiration rate. For Earth, the transition from water-limited to energy-limited begins to occur at approximately A = 0.39, the threshold for humid climates and the point at which soil moisture reaches its field capacity (the maximum water content of a wet soil held through capillary forces with negligible runoff). Below this threshold, A and soil moisture are well-correlated with each other, while above this threshold, soil moisture approaches its upper limit and increases only weakly with A. For some of the more exotic climates we explore in this paper, the correspondence between A and soil moisture is not as close for reasons unique to warm diurnally-driven circulations; we will return to this point in the Discussion section. 

To assess the possibility of ``superhabitable" planets, we examine several complementary aridity-based metrics for the different planets in our ensemble.  The simplest is the mean and spatial distribution of aridity itself.   Another is the fraction of the total land area occupied by wet vs. dry climates. We use this to create a regional habitability metric defined as humid aridity values at above-freezing mean temperatures. Other metrics are also possible, e.g., to quantify the fraction of time over an orbit during which a grid point contains liquid water under habitable temperature conditions, but we do not pursue this in this paper.

\section{Mechanisms controlling land habitability}

For different rotation periods (P), climates of Earth-like planets occupy three different dynamical regimes that reflect different dominant modes of dynamical heat transport, as discussed in our Part I and Part II papers: 

(1) A quasi-geostrophic regime (P $<$ 8 d) in which a tropical Hadley cell produces rising motion (promoting precipitation) near the equator and subsidence (suppressing precipitation) at higher latitudes. The latitudinal extent of the Hadley cell is limited by synoptic scale eddies driven by latitudinal temperature gradients via baroclinic instability at higher latitudes.  The eddies create a secondary maximum of rainfall in the extratropics.

(2) A quasi-barotropic regime (8 $<$ P $<$ 32 d) that sets in when the Rossby radius of deformation (approximately the spatial scale of baroclinic eddies) approaches and then exceeds the size of the planet. By this point the Hadley cell has expanded to cover the full latitudinal range from equator to pole, the baroclinic eddies have disappeared, and temperatures have become relatively uniform across the planet.

(3) A diurnally driven regime (P $>$ 32 d) in which the rotation period becomes comparable to or greater than the radiative relaxation time of the atmosphere and the circulation switches from primarily equator-pole heat transport to primarily day-night transport.  Rising motions (and thus precipitation and high temperatures) follow the progression of the Sun, and nightside temperatures cool to an extent consistent with the heat capacity of the surface and atmosphere.

\begin{figure} [ht]
\includegraphics[scale=0.47]{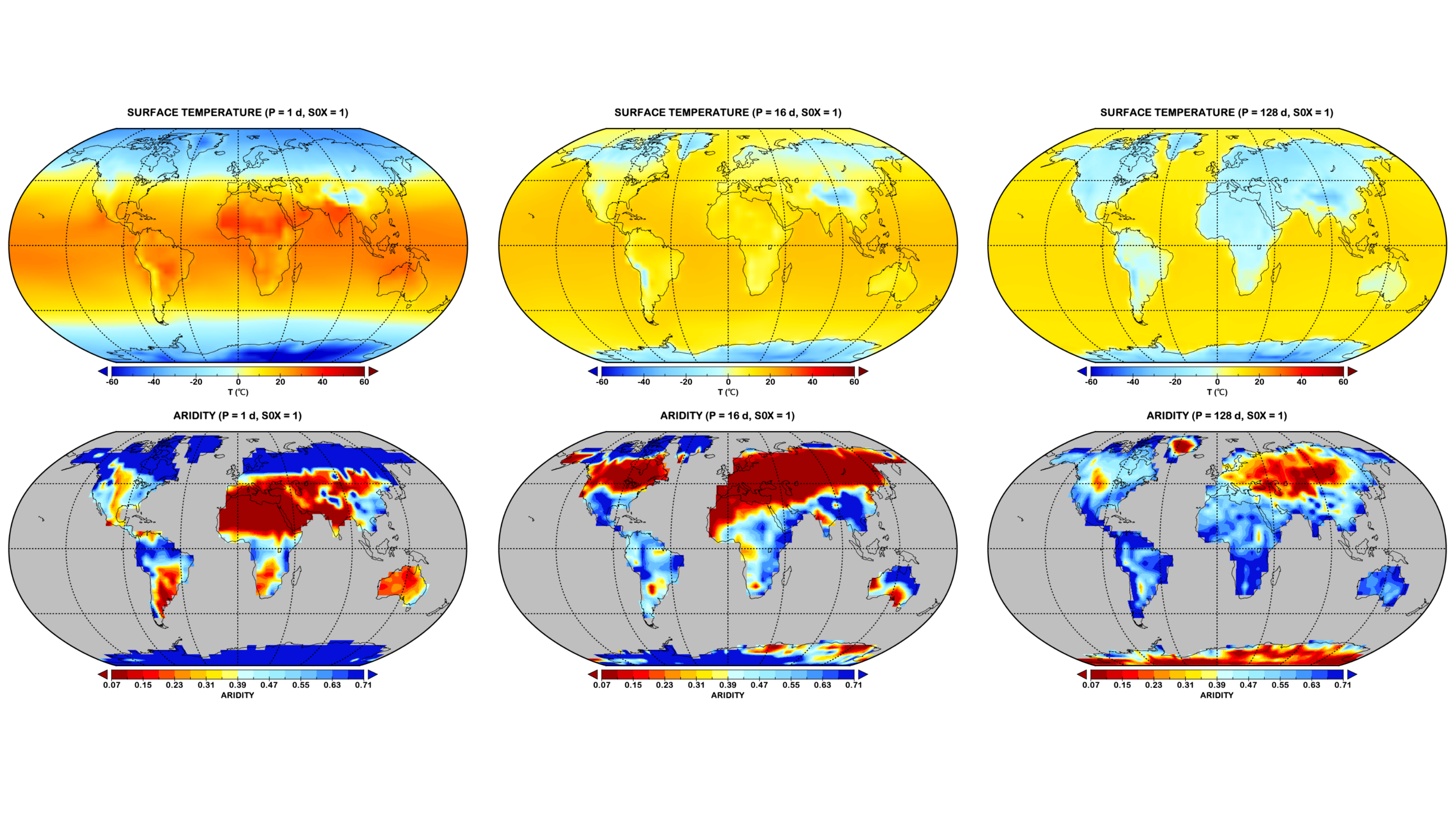}
\centering
\caption{\small Annual mean surface temperature (upper panels) and aridity index (lower panels) for S0X = 1 for planets with P = (left) 1 d, (middle) 16 d, and (right) 128 d.}
\label{fig1}
\end{figure}

Figure \ref{fig1} illustrates the resulting annual mean surface temperature and aridity patterns for planets illuminated with modern Earth's solar constant (S0X = 1) for rotation periods representative of the three dynamical regimes.  For P = 1 d, the equator-pole temperature gradient is large.  Two latitude bands of humid climate conditions (A $>$ 0.39) exist, one straddling the equator and another poleward of $45^\circ$ latitude in both hemispheres, in addition to some coastal regions with humid conditions.  These correspond to the rising branch of the Hadley cell and the midlatitude storm tracks, respectively. A is actually largest at high latitudes, but these regions are below freezing and snow/ice-covered and thus not habitable by the standard definition used for exoplanets.  (This differs from actual modern Earth, with its non-zero obliquity and thus seasons, which allow some areas of the polar regions to stay above freezing for part of the year.)  Hyper-arid desert areas occupy subtropical land areas not too different from those on modern Earth.

At P = 16 d, temperatures are moderately warm everywhere because of the planet-wide Hadley cell.  The only exceptions are local regions of high topography and land areas poleward of $60^\circ$ latitude in both hemispheres, which are slightly below freezing.  Humid conditions in the now broad rising branch of the Hadley cell span most of the land area equatorward of $30^\circ$ latitude.  Hyper-arid deserts are shifted poleward and broadened to include much of Eurasia and Canada.  At P = 128 d, ocean temperatures are a bit cooler than at 16 d but similarly uniform over the planet.  Land temperatures, though, are much cooler - slightly below freezing in most places.  This occurs because of the strong diurnal cycle of solar heating and the small thermal inertia of land relative to ocean:  Temperatures rise dramatically during the day, limited somewhat due to shielding by thick clouds that form in rising regions \citep{Yang2014}, and then cool sharply during the long night when clouds clear.  The resulting aridity pattern exhibits humid conditions at all latitudes except for some continental interior regions.  Although the diurnal mean temperatures over land are below freezing and thus excluded by our temperature constraint as being habitable, these regions actually are above freezing for part of the day with rain falling.

\begin{figure}[ht]
\includegraphics[scale=0.47]{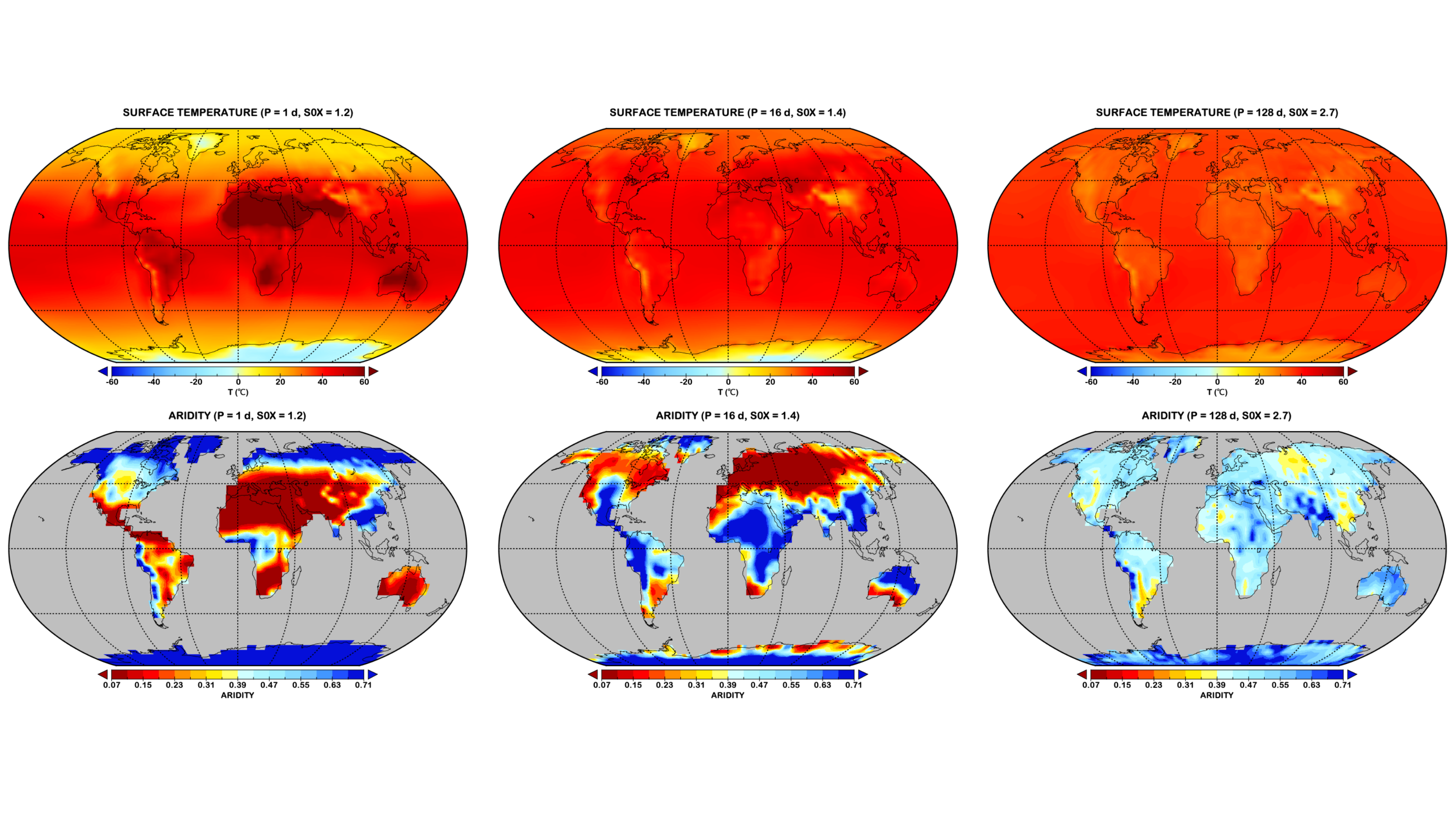}
\centering
\caption{\small As in Fig. 1, but for the highest value of S0X simulated for each rotation period.}
\label{fig2}
\end{figure}

Figure \ref{fig2} shows the surface temperature and aridity distributions for the same three rotation periods but for the highest insolation values for which the model reached equilibrium.  Surface temperatures are now above freezing everywhere for P = 1 d and 16 d except in the interior of Antarctica and Greenland, and above freezing everywhere for P = 128 d.  For P = 1 d temperatures are intolerably hot by human standards in the tropics but still within the tolerances for complex life.  For the longer rotation periods, even tropical temperatures are no worse than the upper limits that occur on modern Earth.  The aridity patterns remain fairly similar to those for S0X =1 for P = 1 d and 16 d except for the fact that the very wet polar regions are now mostly above freezing. For P = 128 d, however, the continental interior deserts have vanished, with conditions near the lower end of the humid range almost everywhere except coastal regions, which are somewhat wetter.

\begin{figure}[ht]
\includegraphics[scale=0.4]{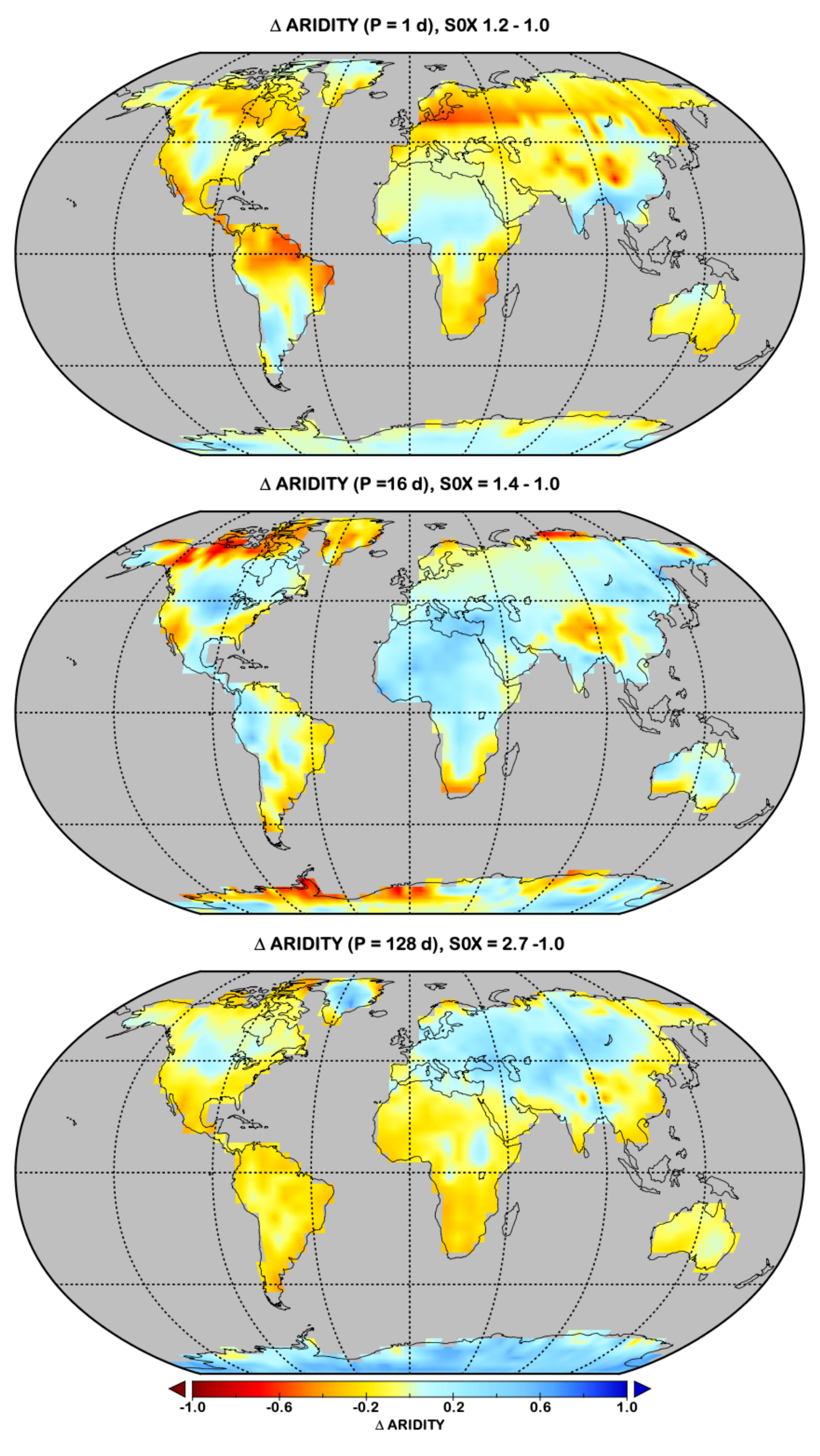}
\centering
\caption{\small Change in aridity with increasing S0X for each rotation period.}
\label{fig3}
\end{figure}

Figure \ref{fig3} shows the aridity changes with S0X for all three rotation periods.  Our simulations are not identical to terrestrial 21st Century climate change experiments, e.g., because our planet has zero obliquity and the change is insolation-driven rather than greenhouse gas-driven.  Solar forcing and greenhouse gas forcing have different spatial distributions, and furthermore many Earth-specific aspects of land dynamics (including the absence of vegetation) are simplified in our simulations.  Nonetheless, we can deduce some features of the aridity response to changing insolation from terrestrial experience.   

The pattern of aridity change we see in the upper panel of Figure 3 is similar in many ways to that predicted for future terrestrial climate change \citep{Scheff2017}   - specifically, a drying of the subtropics and parts of the midlatitudes caused by poleward expansion of the downwelling branch of the Hadley cell with warming and an accompanying shift in the storm track \citep{Seager2007}, with some regional wetting in parts of the tropics where precipitation increases more than evaporation.

The situation for P = 16 d (middle panel) is different, because for this rotation period the Hadley cell already extends almost to the pole and there are no baroclinic eddies in the middle/high latitudes to influence its strength or extent.  In this limit of slow rotation (i.e., large Rossby number), the Hadley cell is controlled purely energetically \citep{Schneider2010}.  In response to warming, water vapor increases at a rate determined by the Clausius-Clapeyron equation, but radiative cooling increases more slowly, implying that the convective mass flux and thus the strength of the Hadley cell weaken \citep{HeldSoden2006}.  This is the case for our P = 16 d experiments, with the peak streamfunction of the mean meridional circulation only 60\% as strong for S0X = 1.4 as it is for S0X = 1.  Thus at low latitudes, the aridity index increases because of the precipitation increase associated with the warming, while in the descending branch of the Hadley cell, subsidence drying weakens with warming, preventing a significant evaporation increase.

At P = 128 d, although a day-night circulation component now exists, there still remains a weaker (75\%  as strong as at P = 16 d) mean meridional circulation and weak Intertropical Convergence Zone when S0X = 1, with subsidence and dry conditions especially over the interiors of midlatitude continents.  At S0X = 2.7, the mean meridional circulation has almost completely disappeared. Thus the tropics dry out and the midlatitude continental subsidence areas moisten. 

\begin{figure}[ht]
\includegraphics[scale=0.485]{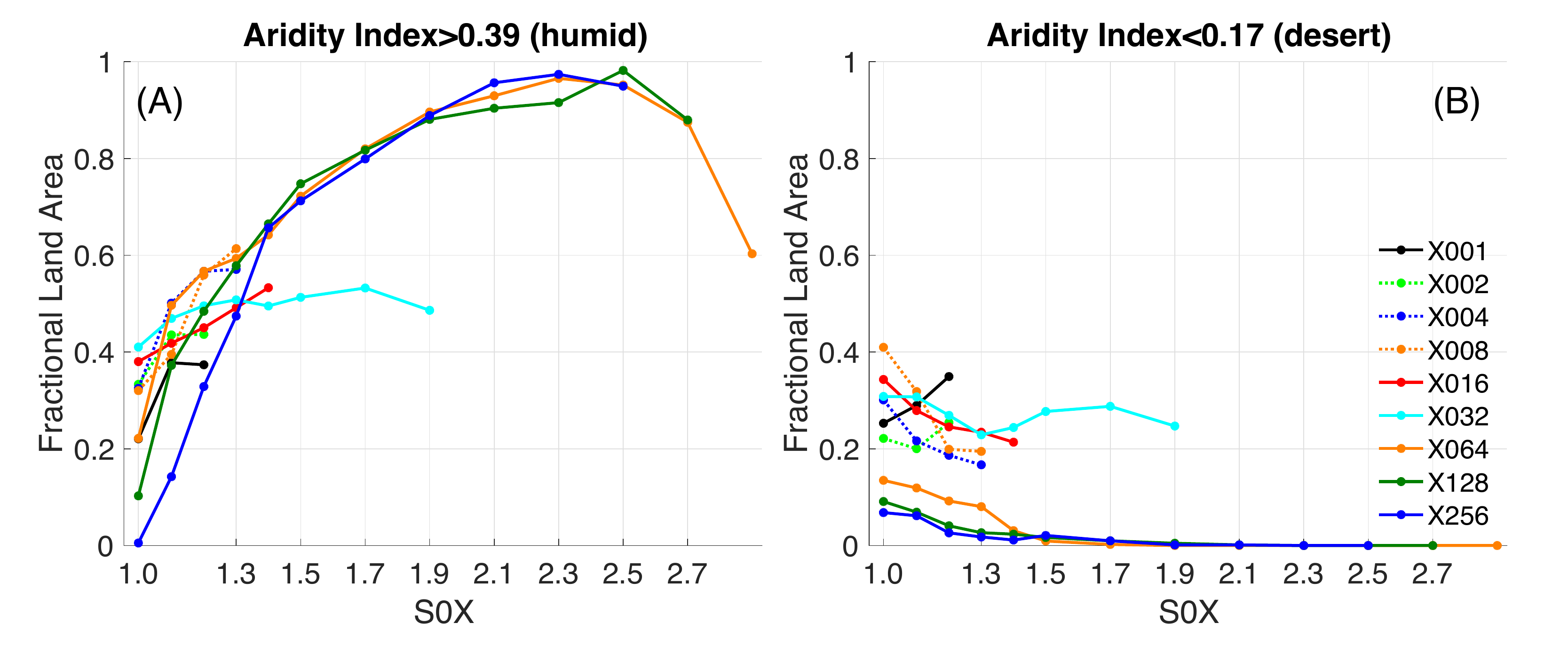}
\caption{\small Left (a): Fraction of land area with humid climate aridity index values and with surface temperatures above 0$^\circ$C as a function of insolation and rotation. Right (b): As in (a) but for desert climate aridity index values.  The legend in the right panel identifies the rotation periods in Earth days.}
\label{fig4}
\end{figure}

Figure \ref{fig4} shows how the ``humid" and ``desert" land area fractions vary with rotation and insolation.  We use this as a zeroth-order index of the extent of habitability, assuming that even though life exists in the world's deserts, detectable land surface life on exoplanets will be restricted to areas with more abundant vegetation. Choosing a more restrictive value of A, e.g., A $<$ 0.05 (not shown), the threshold for hyper-arid desert climates on Earth, has little effect on the rotation and insolation dependence in Figure \ref{fig4}. 

At modern Earth insolation, land habitability (defined by humid aridity values + above-freezing temperatures) is limited primarily by temperature. At the more rapid rotation rates, low- and mid-latitude land is habitable except in the desert regions, while at high latitudes temperatures below freezing limit habitability. At very slow rotation and modern Earth insolation, mean temperatures are below freezing over all land areas due to nighttime cooling and are thus excluded as habitable, although some of these areas may be above freezing during daytime. Higher insolation leads to above-freezing temperatures up to high latitudes, similar to equable climate periods in Earth's past.  According to the aridity index, these planets are overall more humid and thus have more extensive habitable conditions. The fraction of desert area consistently declines with stellar flux.  The relation to rotation rate is not monotonic:  for modest increases in insolation, humid areas increase in area from sidereal day lengths of 1 d to 8 d,  consistent with the conclusions of our Part II paper. For higher insolation, only the planets with slower rotation rates remain habitable.  For these planets, the habitable fractional land area is almost independent of rotation period, with a peak at ~2.1-2.5 x modern Earth's insolation value.  This rotational independence is due to the dayside shielding of the surface by clouds that sets in at long rotation periods, an effect seen in multiple models \citep{Yang2014,Way2018}.  

\section{Biomes, superhabitability, and un-Earthlike climates}

On Earth, life can exist under just about any conditions found on the surface.  Detecting life from great distances on an exoplanet surface, though, favors areas of the surface with abundant vegetation, hence our emphasis on climates with high values of the aridity index (Fig. 4).  Traditional biome classifications \citep{Holdridge1967,Whittaker1975,Koppen2011,RK2011} typically differentiate biomes based on observed associations with annual or seasonal precipitation and temperature.  For example, warm and wet climates may be dominated on Earth by tropical rain forests, moderately wet climates by forests, semi-arid climates by savanna and shrubland, and arid climates by desert or grassland.  

Our ``humid" climate aridity index threshold roughly identifies the first two classes of biome listed above, but our planets contain regional climates typical of all the known Earth biomes. Figure 4 puts some of the ideas about ``superhabitable planets" advanced by \citet{HA2014} into a quantitative perspective.  It suggests that planets illuminated slightly more strongly and rotating somewhat more slowly than modern Earth are moderately ``superhabitable" because more of their land surface (and especially more of the area that receives abundant rain) is above freezing. This is consistent with the conclusion reached in Part II.  Figure 4 also suggests, though, that very slowly rotating planets at high insolation can have close to 100\% of their land surface with high aridity values, which would imply that these are the ultimate superhabitable planets.

In general, changes in the aridity index with rotation and insolation (Figs. 1, 2) can be interpreted as changes in the spatial distribution and fractional coverage of existing terrestrial biome types that might exist on another planet and can be explained by basic tenets of meteorology. \citet{HA2014}, for example, proposed that smaller continents would be more conducive to superhabitability because more of their surface would be moderated by ocean influences with less interior desert area. This effect is seen in the lower right panel of Figure 1:  The large Eurasian continent has an extensive interior arid or hyper-arid region for slow rotation, while the somewhat smaller North American continent has only a small interior desert and the even smaller Australian continent has a humid climate throughout.

This interpretation of the aridity index assumes, though, that it applies in the same way to other planets as it does on Earth.  On Earth atmospheric water fluxes are generally good predictors of conditions beneath the surface.  For example, hot and humid climates produce copious rain, moderate evaporation (moderate to large A) and large subsurface soil water reservoirs that support tropical rainforests because they are not water-limited; hot and dry climates (small A) produce little rain and small below-ground water reservoirs and are thus deserts because they are water-limited; moderate climates (mid-range A) with varying precipitation, temperature, and subsurface water are populated with temperate forests that can be energy- or water-limited; and cooler wetter climates have moderate precipitation but little evaporation and are thus covered by energy-limited boreal forests \citep{Koster2009}.   

The correspondence between the flux and reservoir views breaks down, however, at high latitudes on strongly irradiated very slowly rotating planets. One such example is our P = 128 d, S0X = 2.7 case (right panels, Fig. 2). This planet has a moderately humid aridity index and very warm temperatures virtually everywhere, suggesting a planet-wide tropical rainforest-like climate.  Actually, though, this planet has relatively little subsurface water poleward of 45$^\circ$, inconsistent with a forested climate of any kind by terrestrial standards.

\begin{figure}[!htb]
\includegraphics[scale=0.485]{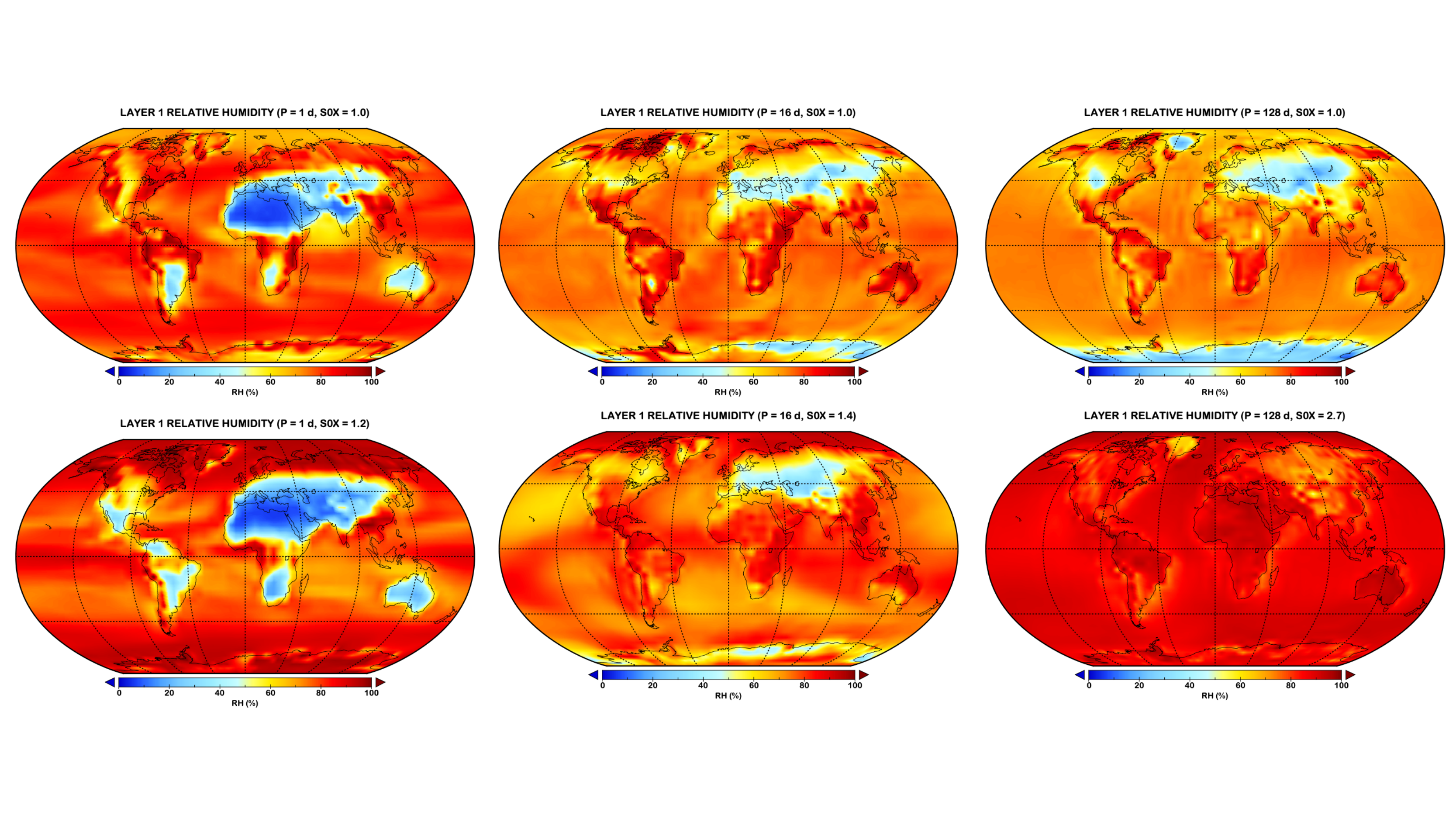}
\centering
\caption{\small Relative humidity in the lowest atmospheric model layer for rotation periods 1, 16, and 128 d for S0X = 1 and for the highest S0X with an equilibrated climate for each period.}
\label{fig5}
\end{figure}

\begin{figure}[!htb]
\includegraphics[scale=0.485]{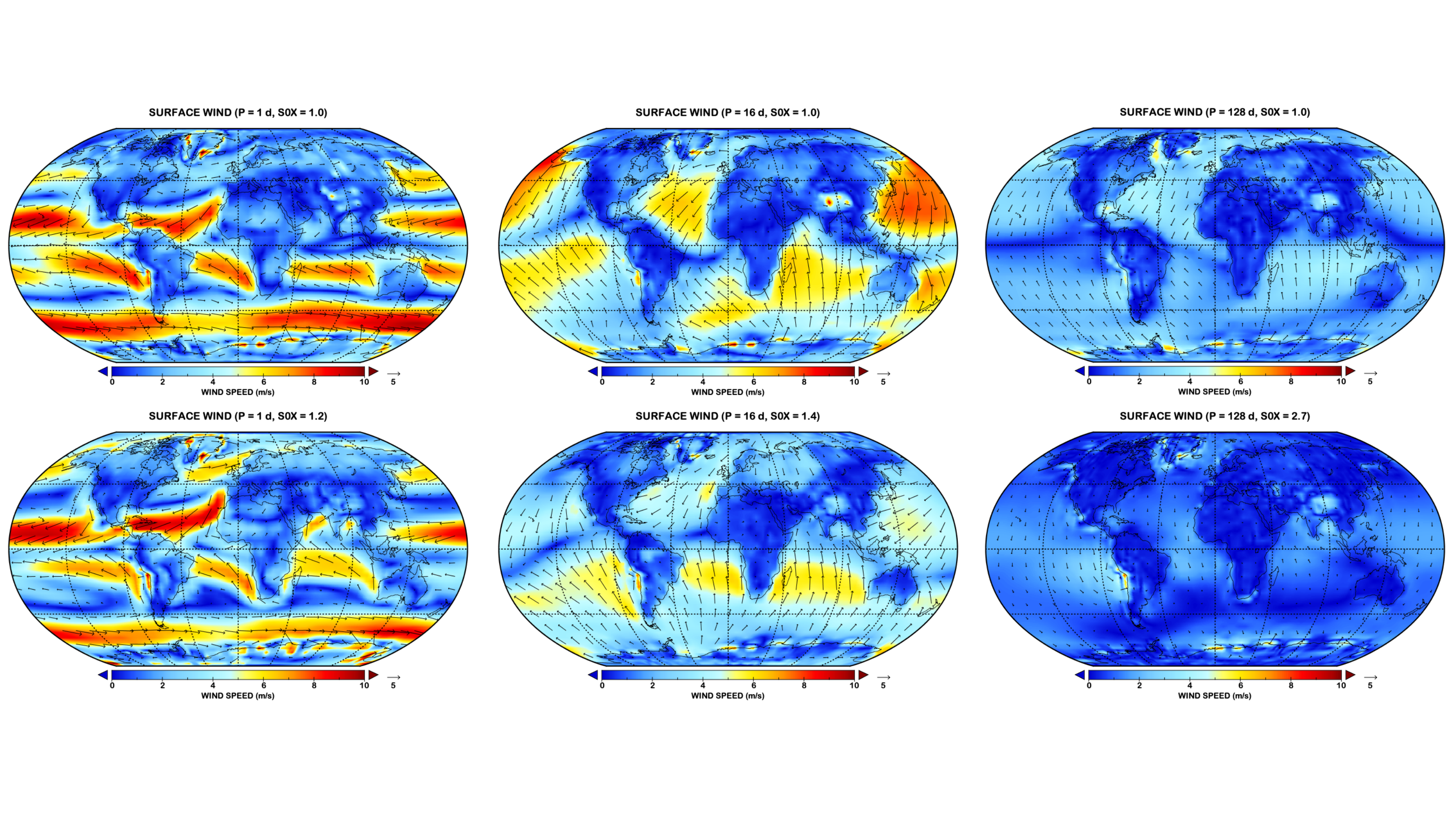}
\centering
\caption{\small Surface wind speed and direction for rotation periods 1, 16, and 128 d for S0X = 1 and for the highest S0X with an equilibrated climate for each period.}
\label{fig6}
\end{figure}

To understand why, Figures 5 and 6 show the relative humidity (RH) in the lowest model layer and the surface wind speed, respectively, for the 6 planets shown in Figures 1 and 2. These quantities are the most important determining factors for PET.  Typically low-level RH is an indicator of the atmospheric general circulation: High values occur in regions of moisture convergence, rising motion, and precipitation, and low values prevail in regions of subsiding motion that brings dry air down to the surface from higher altitudes, producing large PET.  Signatures of this can be seen in Figure 5 for 5 of our 6 planets, with relatively high RH over rainy tropical continents and lower RH over dry subtropical or midlatitude continents, depending on the rotation period.  For the P = 128 d, S0X = 2.7 planet, though, low-level RH is extremely high ($>$80\%) almost everywhere on the planet, including at high latitudes. This suppresses PET, which depends on the humidity difference between the ground and the overlying air.  Likewise, surface wind, which controls the vapor flux from surface to air for a given RH, generally weakens with increasing insolation and slowing rotation, with especially weak winds planet-wide at P = 128 d and S0X = 2.7. Both the higher RH and the weak wind speeds for this planet reduce PET, creating larger values of the aridity index than occur in other experiments at high latitudes despite the small amount of water stored below ground.  Unlike at low latitudes, which are rainy during the daytime due to the onset of a day-night general circulation component for very slowly rotating planets, high RH at high latitudes is not accompanied by rain because of the absence of strong rising motion.  Consequently, high latitudes maintain a warm and extremely humid yet dry climate, a regime with no counterpart on Earth.

Table 1 shows metrics that relate the atmospheric water cycle to the energy cycle to highlight the unique character of the very slow rotation, high insolation climate.  The precipitable water (PW), defined as the vertically integrated water vapor content of the atmosphere (in kg/m$^2$), is equivalent to the depth of a liquid water layer (in mm) that would form on the surface if all water vapor were condensed and precipitated.  Unsurprisingly, it increases dramatically with S0X due to the temperature dependence of the saturation humidity, and is a weaker function of rotation.  The precipitation Pr varies less as S0X increases and even decreases for the 128 d rotation period case, a consequence of rain being restricted to the daytime tropics.  The ratio of these reservoir and flux quantities is the atmospheric water residence time

t$_{res}$ = PW/Pr. 

The equivalent metric for the global energy cycle, the ratio of the atmospheric heat content to the rate of thermal emission to space F is the radiative relaxation time

t$_{rad}$ = pc$_{p}$T/gF

where p is the atmospheric pressure (Pa), c$_{p}$ the specific heat at constant pressure (J/K-kg), T the temperature (K), g the acceleration of gravity (m/s${^2}$), and F the emitted thermal flux to space (W/m${^2}$).

\begin{center}
\begin{table}[ht!]
\caption{Water and energy cycle quantities for experiments in Figure 5}
 \begin{tabular}{l c c c c c l}
 \hline
 Rotation period (d) & S0X & PW (mm) & Pr (mm/d) & t$_{res}$ (d) & t$_{rad}$ (d) & t$_{res}$/t$_{rad}$ \\ 
 \hline\hline
 1 & 1.0 & 22.1 & 3.0 & 7.4 & 65.9 & 0.11 \\
 \hline
 1 & 1.2 & 108.5 & 4.6 & 23.6 & 63.0 & 0.37 \\
 \hline
 16 & 1.0 & 15.0 & 3.1 & 4.8 & 62.3 & 0.08 \\
 \hline
 16 & 1.4 & 124.9 & 5.3 & 23.6 & 58.8 & 0.40 \\
 \hline
 128 & 1.0 & 11.8 & 2.9 & 4.1 & 65.0 & 0.06 \\
 \hline
 128 & 2.7 & 185.9 & 2.4 & 77.5 & 65.3 & 1.19 \\ 
 \hline
\end{tabular}
\end{table}
\end{center}

These time scales indicate the extent to which phase changes of water vs. radiative heating/cooling drive the general circulation, indicated in Table 1 by the ratio t$_{res}$/t$_{rad}$.  For modern Earth (and for our P = 1 d, S0X = 1 planet, which is similar to modern Earth), the water residence time is about a week, much shorter than the 2 month radiative relaxation time.  Thus for example the Hadley cell, while owing its existence to the equator-pole insolation gradient, is heavily influenced by latent heat release in deep convective precipitating storms in its rising equatorial branch.  Table 1 shows that t$_{res}$/t$_{rad}$ remains very small regardless of rotation rate for S0X = 1.  The contrast in time scales is less dramatic for the P = 1 d, S0X = 1.2 and the P = 16 d, S0X = 1.4 cases, and thus the atmosphere retains more water vapor in these cases, but water still cycles through the atmosphere more quickly than radiative processes heat and cool it.  The P = 128 d, S0X = 2.7 case is the exception, with t$_{res}$ slightly longer than t$_{rad}$, implying that the circulation is more strongly controlled by radiative heating-cooling gradients than for the other planets.  On this planet, water's role is more passive, the atmosphere serving primarily as a second reservoir that stores a significant fraction of the water, especially at high latitudes, that is ordinarily stored beneath the planet surface.  This is unlike any biome found on Earth. It is perhaps closest to a tropical savanna climate, but in our case the brief dayside ``rainy" season that occurs poleward of ~45 degrees latitude delivers only 1-2 mm/d of rain, with 0.5 mm/d or less the rest of the year, even though humidity is oppressively high most of the time.  This occurs despite the atmosphere having almost an order of magnitude more water vapor than modern Earth on average, and about 3 times more than the most humid, rainy parts of Earth's tropics - a variant on the ``water, water everywhere, nor any drop to drink" lament of the sailor stuck in the doldrums in \textit{The Rime of the Ancient Mariner}.  

\section{Discussion}
\subsection{Limitations}\label{sec:Limitations}

This paper utilizes a limited ensemble of Earth-like planets in which only two variables (insolation and rotation) are varied.  The ensemble has several fixed properties which, if relaxed, would be expected to have significant effects on the fraction of land that is habitable.  First, our planets have no seasonality, and thus unlike Earth, where the land surface at some latitudes is warmer than 0$^\circ$C for only part of the year, in our ensemble the land surface oscillates from above to below freezing over the diurnal cycle for the more slowly rotating planets.  For these planets, whose land surfaces experience extended periods with no insolation and thus mean temperatures below freezing despite limited daytime warm intervals, it is not clear how habitability should be characterized. 

Second, our ensemble is for planets with an Earth-like land-ocean distribution.  This influences the insolation and rotation dependence of the habitable fraction of land (Fig. 4) primarily because the latitude of the descending branch of the Hadley cell, and thus the locations with A $<<$ 0.39, shift poleward as rotation period increases from 1 d to 32 d.  A planet with a radically different distribution of land, e.g., with primarily equatorial vs. primarily polar continents, would most likely have a larger/smaller fraction of land area with humid and above-freezing climates than the planets in our ensemble.  A more fundamental limitation of the ensemble is that all our planets are dominated by oceans (i.e., large, deep unbounded water masses whose volume changes negligibly as climate changes). For such planets the surface water distribution partly determines the climate, i.e., dry climates occur where land is far from any ocean source of moisture such as the interior of Eurasia.  One can imagine another type of planet dominated by land, with no oceans but inland lakes and small seas whose existence, location and size depend on the extraction of water from the soil and delivery by the atmospheric circulation to other parts of the planet. (Titan is such a planet, but with methane playing the role of water.)  For such planets the climate determines the surface water distribution instead, and the size of the soil moisture reservoir is then a key metric for regional habitability.  This type of planet will be the subject of a future paper.

Third, our ensemble contains only ``warm" planets (as highly or more highly irradiated than modern Earth).  Of course planets that receive moderately less insolation than modern Earth can be habitable too.  For a modern Earth-like atmospheric composition, we expect such planets to have less habitable land area than modern Earth because of their greater snow/sea ice coverage, similar to partial or total snowball periods in Earth's past (e.g., \cite{SohlChandler2007}).  However ``equable" climate periods with moderate temperatures all the way to the poles have occurred several times in Earth's past when greenhouse gas concentrations were elevated despite a slightly dimmer Sun \citep{Burke2018}, and these may have habitable fractions more like our S0X $>$ 1 rapid rotators.

Finally, our emphasis in this paper is on the use of a proxy that implies the influence of climate on the spatial distribution and character of complex life.  Our land surface, though, consists of bare soil.  On an actual vegetated surface life will modulate the climate via its effect on surface albedo and the transpiration rate of water. These effects will be left to a future study.

\subsection{Effects on Surface Biosignature Detection}\label{Biosignature Detection}

The prime motivation for exploring external parameter settings that are conducive to extensive habitable areas on a planet is to inform thinking about future biosignature detection.  Biosignatures include atmospheric gases that are the products of Earth-like life and direct indications of life on the land surface of a planet from the spectral dependence of reflected starlight \citep{Kiang2007,Schwieterman2018}.  On Earth, land and ocean contribute roughly equally to the net primary production \citep{Woodward2007}.  On land, forests (which inhabit ``humid" aridity index regions) make up the bulk of the biomass \citep{Baron2018,Erb2018}, but the relationship between the production of biosignature gases and biomass amount is complex \citep{Seager2013,Schwieterman2018}.  Our simulations are more directly relevant to surface-based biosignature detection, but have implications for production of biogenic gases that may be relevant to particular climate types and whose potential for detectability in different environments remains to be explored.  For example, it is hypothesized that some tree types emit isoprene in response to heat stress \citep{Sharkey2008};  while of low detectablity on Earth, we cannot rule out the detectabilty of a surface-derived biogenic gas with a similar function on an exoplanet.

All things being equal, a planet with a larger habitable area, and especially a larger area of the abundant vegetation that we assume to occur in humid, rainy climates, should be most conducive to surface biosignature detection. Clouds, though, are also correlated with humid, rainy climates.  Thus the question arises whether clear skies occur frequently enough on such planets to permit detection of surface indicators of life such as the vegetation red edge, or whether paradoxically a planet with less ideal climate conditions might be more optimal for detection because of visibility constraints.

Figure 7 represents a first attempt to anticipate the effects of instellation and rotation on remote sensing of the surface by a future direct imaging mission.  It shows the fraction of starlight incident on the planet that reaches the surface (whether direct or diffuse) for the three rotation periods and two insolation values explored in previous figures.  This provides only an upper limit on detectability, since some starlight that reaches the surface is absorbed there or by the atmosphere as it travels back up, or is reflected back down by clouds.  In general, ``blue skies" (large fractions of incident starlight that reach the surface) decrease as insolation increases. Combined with the behavior seen in Figure 4, this implies that an increase in habitable area is offset by a decrease in visibility of the surface, because the greater upward transport of water vapor evaporated or transpired from the surface as stellar heating increases results in more and/or more opaque clouds.  The rotation dependence is not monotonic:  As rotation period increases from modern Earth's value and the Hadley cell broadens and expands poleward, the tropical band of convective storms widens and at least partly obscures more area, but this is more than offset by the broad regions of subsidence that suppress clouds over most of the rest of the planet and the disappearance of the cloudy midlatitude storm tracks.  At even slower rotation, as the general circulation makes a transition to primarily day-night, visibility of the surface begins to decrease again as optically thick clouds begin to congregate on the dayside (where the incident starlight needed to image the surface is).  Thus, planets that rotate somewhat more slowly than Earth, with insolation not much stronger than Earth receives, appear to be the best candidates for surface detection of the vegetation red edge (as well as ancillary habitability indicators such as ocean glint).

We note several caveats.  Figure 7 represents only the time mean state, so even regions with partial visibility of the surface in the time mean can be expected to have clear skies sometimes as storms come and go.  Also, our ensemble consists only of planets with zero obliquity and thus no seasons.  On Earth, seasonality manifests itself in the tropics as primarily wet/cloudy vs. dry/clear times of year. In fact, tropical vegetation is most sensitive to available sunlight in these regions: Net primary production is greatest in rain forests in the dry season when they receive the most sunlight, a fact deduced from remote sensing of the red edge by Earth-orbiting satellites \citep{Seddon2016}.  Thus, we expect detection of the red edge to be more feasible for planets with modest nonzero obliquities that allow densely vegetated regions to be visible remotely during at least part of the year without radically changing the mean climate from the simulations examined in this paper.

\begin{figure}[ht]
\includegraphics[scale=0.48]{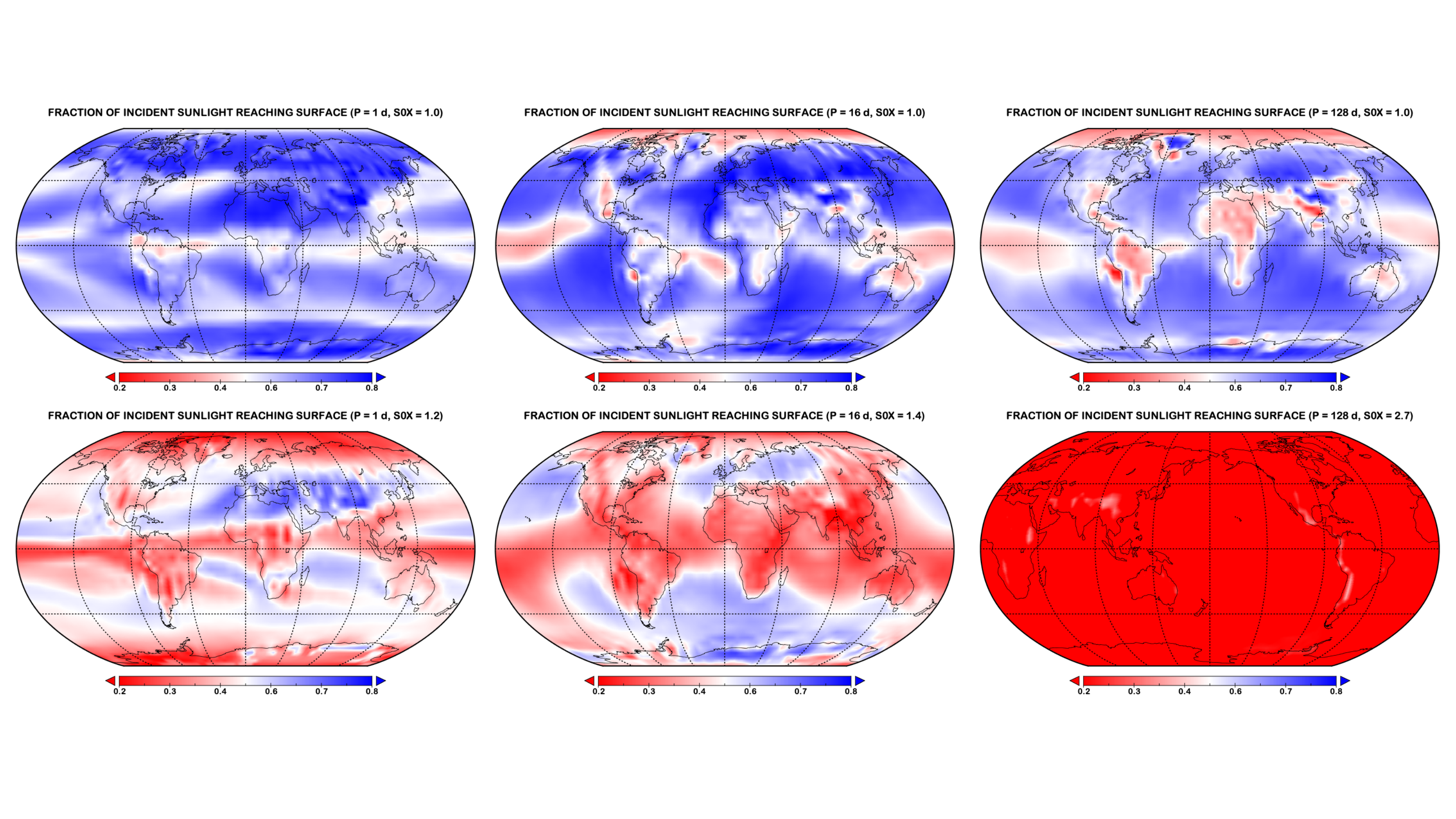}
\centering
\caption{\small Fraction of incident starlight reaching the surface for low and high instellation values and for rotation periods of 1, 16, and 128 d.}
\label{fig7}
\end{figure}

\section{Conclusions}

The presence of surface liquid water is considered central to the question of whether a rocky exoplanet is a good candidate to host life.  For land-based, complex life, detectability is paramount, yet the spatial distribution of surface water has for the most part been ignored to date in exoplanet studies.  We have shown that a simple metric of the competition between the supply of water to the surface and the atmosphere's demand of water from the surface is predictable from external parameters that are either already known (insolation) or may be constrained by future observations (rotation).  In most cases this metric is a good proxy for the abundance of liquid water beneath the surface, as long as it is restricted to regions with temperatures above the freezing point.  The exception is highly irradiated, very slowly rotating planets whose higher latitudes are drier than the aridity index predicts, because such planets' near-surface atmospheres are almost uniformly very humid with weak winds, which distorts the relationship between water fluxes into and out of the surface and the storage of water beneath the surface.

We find that in general, the fraction of an Earth-like planet with humid conditions and thus the potential for abundant land surface life increases as insolation increases and is a maximum at intermediate rotation periods for which the Hadley cell extends to the poles but a significant day-night circulation does not exist.  This is roughly consistent with the temperature-only estimate of fractional habitability derived in Part II.  Consequently, ``superhabitable" planets should be expected to be found once the search for habitable planets is better able to sample regions outside the tidal locking radius of stars yet close enough for incident starlight to support water in its liquid phase on the surface.

Despite the tendency for warmer, more slowly rotating planets to be more favorable to extensive abundant surface life, the presence of more cloudy conditions on such planets compromises remote detectability of such life by a future direct imaging mission.  This conclusion has been derived from mean conditions, however.  In the future, a more reliable estimate might be obtained using high temporal frequency model results to allow the frequency of occurrence of clear skies to be more directly estimated, which in turn would help define the amount of observing time required to successfully sense land surface vegetation or ocean glint.  Since clouds are a highly uncertain aspect of 3-D models, it would be useful for such analyses to be conducted with multiple independent 3-D models.  Such studies should be a prelude to any serious attempt to define the specifications of any future direct imaging mission, since estimates of surface detection to date have been conducted only with Earth analog planets.

\acknowledgements
This work was supported by the NASA Astrobiology Program through collaborations arising from our participation in the Nexus for Exoplanet System Science, by the NASA Planetary Atmospheres Program, and by the GSFC Sellers Exoplanet Environments Collaboration (SEEC). Computing resources for this work were provided by the NASA High-End Computing (HEC) Program through the NASA Center for Climate Simulation (NCCS) at Goddard Space Flight Center http://www.nccs.nasa.gov. 
M.J.W. and N.Y.K. acknowledge support from the
GSFC Sellers Exoplanet Environments Collaboration (SEEC),
which is funded by the NASA Planetary Science Division
Internal Scientist Funding Model.

\bibliographystyle{apj}
\bibliography{bibliography}

\begin{thebibliography}{}
\expandafter\ifx\csname natexlab\endcsname\relax\def\natexlab#1{#1}\fi
\providecommand{\url}[1]{\href{#1}{#1}}

\bibitem[{{Abbot} {et~al.}(2012){Abbot}, {Cowan}, \& {Ciesla}}]{Abbot2012}
{Abbot}, D.~S., {Cowan}, N.~B., \& {Ciesla}, F.~J. 2012, \apj, 756, 178

\bibitem[{{Abe} \& {Abe-Ouchi}(2003)}]{Abe2003}
{Abe}, Y., \& {Abe-Ouchi}, A. 2003, in Lunar and Planetary Science Conference,
  Vol.~34, Lunar and Planetary Science Conference, ed. S.~{Mackwell} \&
  E.~{Stansbery}

\bibitem[{{Abe} {et~al.}(2011){Abe}, {Abe-Ouchi}, {Sleep}, \&
  {Zahnle}}]{Abe2011}
{Abe}, Y., {Abe-Ouchi}, A., {Sleep}, N.~H., \& {Zahnle}, K.~J. 2011, AsBio, 11,
  443

\bibitem[{Allen {et~al.}(1998)Allen, Pereira, Raes, \& Smith}]{Allen1998}
Allen, R.~G., Pereira, R.~S., Raes, D., \& Smith, M. 1998, Crop
  evapotranspiration-guidelines for computing crop water requirements (Rome:
  Food and Agriculture Organization)

\bibitem[{Allen {et~al.}(2005)Allen, Walter, \& Elliott}]{Allen2005}
Allen, R.~G., Walter, I.~A., \& Elliott, R. L. e.~a. 2005, The ASCE
  standardized reference evapotranspiration equation (Reston: American Society
  of Civil Engineers).
\newblock \url{https://searchworks.stanford.edu/view/6024791}

\bibitem[{{Arnold}(2008)}]{Arnold2008}
{Arnold}, L. 2008, \ssr, 135, 323

\bibitem[{Bar-On {et~al.}(2018)Bar-On, Phillips, \& Milo}]{Baron2018}
Bar-On, Y.~M., Phillips, R., \& Milo, R. 2018, PNAS, 115, 6506

\bibitem[{Behrenfeld {et~al.}(2005)Behrenfeld, Boss, Siegel, \&
  Shea}]{Behren2005}
Behrenfeld, M.~J., Boss, E., Siegel, D.~A., \& Shea, M.~J. 2005, GBC, 19,
  GB1006

\bibitem[{Burke {et~al.}(2018)Burke, Williams, Chandler, Haywood, Lunt, \&
  Otto-Bliesner}]{Burke2018}
Burke, K.~D., Williams, J.~W., Chandler, M.~A., {et~al.} 2018, PNAS, 115, 13288

\bibitem[{Chandler {et~al.}(2013)Chandler, Sohl, Jonas, Dowsett, \&
  Kelley}]{Chandler2013}
Chandler, M.~A., Sohl, L.~E., Jonas, J.~A., Dowsett, H.~J., \& Kelley, M. 2013,
  GMD, 6, 517

\bibitem[{{Cockell}(2016)}]{CockellMBoC2016}
{Cockell}, C.~S. 2016, Molec. Biol. Cell, 27, 1553

\bibitem[{{Cockell} {et~al.}(2016){Cockell}, {Bush}, {Bryce}, {Direito},
  {Fox-Powell}, {Harrison}, {Lammer}, {Landenmark}, {Martin-Torres},
  {Nicholson}, {Noack}, {O'Malley-James}, {Payler}, {Rushby}, {Samuels},
  {Schwendner}, {Wadsworth}, \& {Zorzano}}]{Cockell2016}
{Cockell}, C.~S., {Bush}, T., {Bryce}, C., {et~al.} 2016, AsBio, 16, 89

\bibitem[{{Colose} {et~al.}(2019){Colose}, {Del Genio}, \& {Way}}]{Colose2019}
{Colose}, C.~M., {Del Genio}, A.~D., \& {Way}, M.~J. 2019, \apj,
  arXiv:1905.09398

\bibitem[{{Cook} {et~al.}(2014){Cook}, {Smerdon}, {Seager}, \&
  {Coats}}]{Cook2014}
{Cook}, B.~I., {Smerdon}, J.~E., {Seager}, R., \& {Coats}, S. 2014, Clim. Dyn.,
  43, 2607

\bibitem[{{Des Marais} {et~al.}(2002){Des Marais}, {Harwit}, {Jucks},
  {Kasting}, {Lin}, {Lunine}, {Schneider}, {Seager}, {Traub}, \&
  {Woolf}}]{DesMarais2002}
{Des Marais}, D.~J., {Harwit}, M.~O., {Jucks}, K.~W., {et~al.} 2002, AsBio, 2,
  153

\bibitem[{{Desch} {et~al.}(2017){Desch}, {Hartnett}, {Kane}, \&
  {Walker}}]{Desch2017}
{Desch}, S.~J., {Hartnett}, H.~E., {Kane}, S.~R., \& {Walker}, S.~I. 2017, in
  Habitable Worlds 2017: A System Science Workshop (Houston: Lunar and
  Planetary Institute)

\bibitem[{Erb {et~al.}(2018)Erb, Kastner, \& Plutzar}]{Erb2018}
Erb, K.~H., Kastner, T., \& Plutzar, C. e.~a. 2018, Nature, 553, 73

\bibitem[{Feng \& Fu(2013)}]{Feng2013}
Feng, S., \& Fu, Q. 2013, ACP, 13, 10081

\bibitem[{Fu \& Feng(2014)}]{Fu2014}
Fu, Q., \& Feng, S. 2014, JGRD, 119, 7863, 2014JD021608

\bibitem[{Fu {et~al.}(2016)Fu, Lin, Huang, Feng, \& Gettelman}]{Fu2016}
Fu, Q., Lin, L., Huang, J., Feng, S., \& Gettelman, A. 2016, JGRD, 121, 2857

\bibitem[{{Held} \& {Soden}(2006)}]{HeldSoden2006}
{Held}, I.~M., \& {Soden}, B.~J. 2006, JCli, 19, 5686

\bibitem[{{Heller} \& {Armstrong}(2014)}]{HA2014}
{Heller}, R., \& {Armstrong}, J. 2014, AsBio, 14, 50

\bibitem[{Holdridge(1967)}]{Holdridge1967}
Holdridge, L.~R. 1967, Life Zone Ecology (San Jose, Costa Rica: Tropical
  Science Center)

\bibitem[{{Jansen} {et~al.}(2019){Jansen}, {Scharf}, {Way}, \& {Del
  Genio}}]{Jansen2019}
{Jansen}, T., {Scharf}, C., {Way}, M., \& {Del Genio}, A. 2019, \apj, 875

\bibitem[{Kaimal \& Finnigan(1994)}]{KF1994}
Kaimal, J.~C., \& Finnigan, J.~J. 1994, Atmospheric Boundary Layer Flows
  (Oxford: Oxford University Press).
\newblock
  \url{https://global.oup.com/academic/product/atmospheric-boundary-layer-flows-9780195062397?cc=us\&lang=en\&}

\bibitem[{{Kane} {et~al.}(2016){Kane}, {Hill}, {Kasting}, {Kopparapu},
  {Quintana}, {Barclay}, {Batalha}, {Borucki}, {Ciardi}, {Haghighipour},
  {Hinkel}, {Kaltenegger}, {Selsis}, \& {Torres}}]{Kane2016}
{Kane}, S.~R., {Hill}, M.~L., {Kasting}, J.~F., {et~al.} 2016, \apj, 830, 1

\bibitem[{Kiang {et~al.}(2007)Kiang, Segura, Tinetti, Govindjee, Blankensip,
  Cohen, Siefert, Crisp, \& Meadows}]{Kiang2007}
Kiang, N.~Y., Segura, A., Tinetti, G., {et~al.} 2007, AsBio, 7, 252

\bibitem[{Kilic {et~al.}(2018)Kilic, Lunkeit, Raible, \& Stocker}]{Kilic2018}
Kilic, C., Lunkeit, F., Raible, C.~C., \& Stocker, T.~F. 2018, \apj, 864

\bibitem[{{Kite} {et~al.}(2017){Kite}, {Sneed}, {Mayer}, \&
  {Wilson}}]{Kite2017}
{Kite}, E.~S., {Sneed}, J., {Mayer}, D.~P., \& {Wilson}, S.~A. 2017, \grl, 44,
  3991

\bibitem[{{Kitzmann} {et~al.}(2015){Kitzmann}, {Alibert}, {Godolt}, {Grenfell},
  {Heng}, {Patzer}, {Rauer}, {Stracke}, \& {von Paris}}]{Kitzmann2015}
{Kitzmann}, D., {Alibert}, Y., {Godolt}, M., {et~al.} 2015, \mnras, 452, 3752

\bibitem[{Kopparapu {et~al.}(2016)Kopparapu, Wolf, Haqq-Misra, Yang, Kasting,
  Meadows, Terrien, \& Mahadevan}]{Kopparapu2016}
Kopparapu, R.~K., Wolf, E.~T., Haqq-Misra, J., {et~al.} 2016, \apj, 819, 84

\bibitem[{K{\"o}ppen(2011)}]{Koppen2011}
K{\"o}ppen, W. 2011, Meteorologische Zeitschrift, 20, 351

\bibitem[{Koster {et~al.}(2009)Koster, Schubert, \& Suarez}]{Koster2009}
Koster, R.~D., Schubert, S.~D., \& Suarez, M.~J. 2009, JCli, 22, 3331

\bibitem[{{Lunt} {et~al.}(2012){Lunt}, {Haywood}, {Schmidt}, {Salzmann},
  {Valdes}, {Dowsett}, \& {Loptson}}]{Lunt2012}
{Lunt}, D.~J., {Haywood}, A.~M., {Schmidt}, G.~A., {et~al.} 2012, EPSL, 321,
  128

\bibitem[{{Middleton} {et~al.}(1997){Middleton}, {Thomas}, \& {United Nations
  Environment Programm}}]{MiddletonThomas1997}
{Middleton}, N.~J., {Thomas}, D., \& {United Nations Environment Programm}.
  1997, World Atlas of Desertification, A Hodder Arnold Publication (Arnold).
\newblock \url{https://books.google.se/books?id=aNqtQgAACAAJ}

\bibitem[{{Mitchell}(2008)}]{Mitchell2008}
{Mitchell}, J.~L. 2008, JGRD, 113, E08015

\bibitem[{{Monta{\~n}{\'e}s-Rodr{\'{\i}}guez}
  {et~al.}(2006){Monta{\~n}{\'e}s-Rodr{\'{\i}}guez}, {Pall{\'e}}, {Goode}, \&
  {Mart{\'{\i}}n-Torres}}]{Montanes2006}
{Monta{\~n}{\'e}s-Rodr{\'{\i}}guez}, P., {Pall{\'e}}, E., {Goode}, P.~R., \&
  {Mart{\'{\i}}n-Torres}, F.~J. 2006, \apj, 651, 544

\bibitem[{Monteith(1975)}]{Monteith1975}
Monteith, J.~L. 1975, Vegetation and the Atmosphere: Principles (Academic
  Press), 278, doi:10.1017/S0014479700008607

\bibitem[{Penman(1948)}]{Penman1948}
Penman, H.~L. 1948, Proc. Roy. Soc. A, 193, 120

\bibitem[{Priestley \& Taylor(1972)}]{PT1972}
Priestley, C. H.~B., \& Taylor, R.~J. 1972, MWR, 100, 81

\bibitem[{{Rodriguez} {et~al.}(2015){Rodriguez}, {Platz}, {Gulick}, {Baker},
  {Fair{\'e}n}, {Kargel}, {Yan}, {Miyamoto}, \& {Glines}}]{Rodriguez2015}
{Rodriguez}, J.~A.~P., {Platz}, T., {Gulick}, V., {et~al.} 2015, \icarus, 257,
  387

\bibitem[{Rubel \& Kottek(2011)}]{RK2011}
Rubel, F., \& Kottek, M. 2011, Meteorologische Zeitschrift, 20, 361

\bibitem[{Scheff \& Frierson(2014)}]{SF2014}
Scheff, J., \& Frierson, D. M.~W. 2014, JCli, 27, 1539

\bibitem[{Scheff \& Frierson(2015)}]{SF2015}
---. 2015, JCli, 28, 5583

\bibitem[{Scheff {et~al.}(2017)Scheff, Seager, Lu, \& Coats}]{Scheff2017}
Scheff, J., Seager, R., Lu, H., \& Coats, S. 2017, JCli, 30, 6593

\bibitem[{{Schneider} {et~al.}(2010){Schneider}, {O'Gorman}, {Schafer}, \&
  {Levine}}]{Schneider2010}
{Schneider}, T., {O'Gorman}, P.~A., {Schafer}, J., \& {Levine}, X.~J. 2010,
  Rev. Geophys., 48, RG3001

\bibitem[{{Schwieterman} {et~al.}(2018){Schwieterman}, {Kiang}, {Parenteau}, \&
  {Harman}}]{Schwieterman2018}
{Schwieterman}, E., {Kiang}, N., {Parenteau}, M., \& {Harman}, C. e.~a. 2018,
  AsBio, 18, 663

\bibitem[{{Schwieterman} {et~al.}(2019){Schwieterman}, {Reinhard}, {Olson},
  {Harman}, \& {Lyons}}]{Schwieterman2019}
{Schwieterman}, E., {Reinhard}, C., {Olson}, S., {Harman}, C., \& {Lyons}, T.
  2019, \apj, 878, doi:10.3847/1538-4357/ab1d52

\bibitem[{{Seager} {et~al.}(2007){Seager}, {Ting}, {Held}, {Kushnir}, {Lu},
  {Vecchi}, {Huang}, {Harnik}, {Leetmaa}, {Lau}, {Li}, {Velez}, \&
  {Naik}}]{Seager2007}
{Seager}, R., {Ting}, M., {Held}, I., {et~al.} 2007, Science, 316, 1181

\bibitem[{{Seager} {et~al.}(2013){Seager}, {Bains}, \& {Hu}}]{Seager2013}
{Seager}, S., {Bains}, W., \& {Hu}, R. 2013, \apj, 775,
  doi:10.1088/0004-637X/775/2/104

\bibitem[{{Seager} {et~al.}(2016){Seager}, {Bains}, \&
  {Petkowski}}]{Seager2016}
{Seager}, S., {Bains}, W., \& {Petkowski}, J.~J. 2016, AsBio, 16, 465

\bibitem[{{Seager} {et~al.}(2005){Seager}, {Turner}, {Schafer}, \&
  {Ford}}]{Seager2005}
{Seager}, S., {Turner}, E.~L., {Schafer}, J., \& {Ford}, E.~B. 2005, AsBio, 5,
  372

\bibitem[{{Seddon} {et~al.}(2016){Seddon}, {Macias-Fauria}, {Long}, {Benz}, \&
  {Willis}}]{Seddon2016}
{Seddon}, A., {Macias-Fauria}, M., {Long}, P., {Benz}, D., \& {Willis}, K.
  2016, Nature, 531, 229

\bibitem[{Sharkey {et~al.}(2008)Sharkey, Wiberley, \& Donohue}]{Sharkey2008}
Sharkey, T.~D., Wiberley, A.~E., \& Donohue, A.~R. 2008, Annals of Botany, 101,
  5

\bibitem[{{Silva} {et~al.}(2017){Silva}, {Vladilo}, {Schulte}, {Murante}, \&
  {Provenzale}}]{Silva2017}
{Silva}, L., {Vladilo}, G., {Schulte}, P.~M., {Murante}, G., \& {Provenzale},
  A. 2017, Int. J. Astrobio., 16, 244

\bibitem[{{Sohl} \& {Chandler}(2007)}]{SohlChandler2007}
{Sohl}, L.~E., \& {Chandler}, M.~A. 2007, Deep-Time Perspectives on Climate
  Change, 61.
\newblock \url{http://geoscienceworld.org/content/9781862396203/9781862396203}

\bibitem[{{Spiegel} {et~al.}(2008){Spiegel}, {Menou}, \&
  {Scharf}}]{Spiegel2008}
{Spiegel}, D.~S., {Menou}, K., \& {Scharf}, C.~A. 2008, \apj, 681, 1609

\bibitem[{Stevenson(2019)}]{Stev2019}
Stevenson, D.~S. 2019, Int. J. Astrobio., doi:10.1017/S1473550419000181

\bibitem[{Thornthwaite(1948)}]{Thornthwaite1948}
Thornthwaite, C.~W. 1948, Geog. Rev., 38, 55

\bibitem[{{Tinetti} {et~al.}(2006){Tinetti}, {Meadows}, {Crisp}, {Kiang},
  {Kahn}, {Fishbein}, {Velusamy}, \& {Turnbull}}]{Tinetti2006}
{Tinetti}, G., {Meadows}, V.~S., {Crisp}, D., {et~al.} 2006, AsBio, 6, 881

\bibitem[{{Villanueva} {et~al.}(2015){Villanueva}, {Mumma}, {Novak},
  {K{\"a}ufl}, {Hartogh}, {Encrenaz}, {Tokunaga}, {Khayat}, \&
  {Smith}}]{Villanueva2015}
{Villanueva}, G.~L., {Mumma}, M.~J., {Novak}, R.~E., {et~al.} 2015, Science,
  348, 218

\bibitem[{{Way} {et~al.}(2018){Way}, {Del Genio}, {Aleinov}, {Clune}, {Kelley},
  \& {Kiang}}]{Way2018}
{Way}, M.~J., {Del Genio}, A.~D., {Aleinov}, I., {et~al.} 2018, ApJS, 239, 22

\bibitem[{{Way} {et~al.}(2016){Way}, {Del Genio}, {Kiang}, {Sohl}, {Grinspoon},
  {Aleinov}, {Kelley}, \& {Clune}}]{Way2016}
{Way}, M.~J., {Del Genio}, A.~D., {Kiang}, N.~Y., {et~al.} 2016, \grl, 43, 8376

\bibitem[{{Way} {et~al.}(2017){Way}, {Aleinov}, {Amundsen}, {Chandler},
  {Clune}, {Del Genio}, {Fujii}, {Kelley}, {Kiang}, {Sohl}, \&
  {Tsigaridis}}]{Way2017}
{Way}, M.~J., {Aleinov}, I., {Amundsen}, D.~S., {et~al.} 2017, \apjs, 231, 12

\bibitem[{Whittaker(1975)}]{Whittaker1975}
Whittaker, R. 1975, Communities and Ecosystems (Macmillan).
\newblock \url{https://books.google.com/books?id=WNgUAQAAIAAJ}

\bibitem[{{Wolf} {et~al.}(2017){Wolf}, {Shields}, {Kopparapu}, {Haqq-Misra}, \&
  {Toon}}]{Wolf2017}
{Wolf}, E.~T., {Shields}, A.~L., {Kopparapu}, R.~K., {Haqq-Misra}, J., \&
  {Toon}, O.~B. 2017, \apj, 837, 107

\bibitem[{Woodward(2007)}]{Woodward2007}
Woodward, F.~I. 2007, Curr. Biol., 17, 269

\bibitem[{{Wordsworth}(2016)}]{Wordsworth2016}
{Wordsworth}, R.~D. 2016, Annual Review of Earth and Planetary Sciences, 44,
  381

\bibitem[{{Yang} {et~al.}(2014{\natexlab{a}}){Yang}, {Bou{\'e}}, {Fabrycky}, \&
  {Abbot}}]{Yang2014}
{Yang}, J., {Bou{\'e}}, G., {Fabrycky}, D.~C., \& {Abbot}, D.~S.
  2014{\natexlab{a}}, \apjl, 787, L2

\bibitem[{{Yang} {et~al.}(2013){Yang}, {Cowan}, \& {Abbot}}]{Yang2013}
{Yang}, J., {Cowan}, N.~B., \& {Abbot}, D.~S. 2013, \apjl, 771, L45

\bibitem[{{Yang} {et~al.}(2014{\natexlab{b}}){Yang}, {Liu}, {Hu}, \&
  {Abbot}}]{YangLiuHuAbbot2014}
{Yang}, J., {Liu}, Y., {Hu}, Y., \& {Abbot}, D.~S. 2014{\natexlab{b}}, \apjl,
  796, L22

\end{thebibliography}

\end{document}